physics/0306021
\documentclass{iopart}

\usepackage[dvipdfm]{graphicx}
\begin{document}

\title[Doughnut beam..]{Mode conversion in optical beam in order to create funnel shaped optical near field to collect cold Rb atoms from MOT }

\author{S. M. Iftiquar}

\address{ERATO Localized Photon Project, Japan Science and Technology, 687-1-17/4F Tsuruma, Machida-shi, Tokyo 194-0004, Japan}

\begin{abstract}
Doughnut shaped light beam has been generated from Gaussian mode ($TEM_{00}$) cw-Ti sapphire laser. After splitting the pump beam into two equal intensity components and introducing unequal convergence and phase delay while they are recombined it results in doughnut mode. Such a beam is tunable and have long  propagation length. The evanescent field generated by 360 mW (at 780 nm wavelength) of such a beam creates optical field of 600 nm decay length with a 5.75 neV repulsive dipole potential. Thus cold Rb atoms (at 10{$\mu$}K or less temperature) released from MOT can be reflected by the surface so that the atoms are collected ultimately at the bottom of the prism. By focussing such doughnut beam with 8 cm focal length converging lens, the dark radius reduces to 22{$\mu$}. We also observe such beam to contain azimuthal phase as well as radial phase distribution.  

\end{abstract}


\maketitle

\section{Introduction}

Localization of tiny particle by doughnut beam was first demonstrated by Ashkin [1]which initiated interest on $TEM_{01}$ laser beam and its application in atom manipulation. Besides, evanescent field atom mirror proposed by Cook and Hill [2] and subsequent interest on atom mirror led to observation of atom reflection [3]. In the same way confinement of cold atom by hollow core laser beam was realized [4]. These two mechanism of atom photon interaction have been nicely manifested through the proposal of atom funnel experiment [5].  Furthermore this hollow core beam ($TEM_{01}$) has drawn increasing attention because of its usefulness to manipulate and guide [6] cold atoms through its dark core, its ability to confine tiny particles [1] as well as atom reflection and collection in atom funnel experiment [5]. With computer generated hologram such a beam has been generated [7]. It is also possible to use cylindrical lens [8], or modified intracavity mirror of laser [9] to generate $TEM_{01}$ mode. Using hollow core optical fiber a doughnut shaped optical near-field was also proposed [10] in order to guide cold atoms. In atom funneling a double cone prism was used to generate cylindrical optical field distribution from a Gaussian beam [5]. The success of the Ashkin trap was the scattering force of photon generated by strong optical field distribution around the particle. However it gradually emerged that the theory of dipole force proposed by Askar'yan [11] can also help to control atoms [12]. For such application doughnut beam can be coupled to a hollow prism container receiving cold atoms so that localized surface waves are generated to collect cold atoms and preserve its state through reflection or, thereafter transport the atoms thought a bottom hole [5].

Various degree of dark-bright contrast and longer effective length of propagation are useful characteristics for its application in atom manipulation. With a presence of non-zero optical intensity at the center a longitudinal scattering force as well as transverse dipole force can effectively transport the atoms. Double cone prism produced beam [5] have steeper optical potential distribution across the bright ring, which is suitable for trapping, than that generated by hologram or other means, however such mode suffers quicker decay with propagation length. Here we report generation of various doughnut shaped beams that covers longer propagation length as well as various degrees of dark bright contrast and analyze its usefulness for atom manipulation.

Cold Rb atoms are generated in a magneto-optical trap (MOT). In order to collect a large number of such cold atoms a thin walled empty inverted pyramidal prism is kept below the MOT center so that after release from the trap the cold atoms fall into the prism under gravitational pull. When light have total internal reflection at glass-air interface an exponentially decaying evanescent field is created [2], a typical range normal to the surface is the optical wavelength of the light used. If frequency of the laser be slightly higher than the resonance absorption line of Rb atoms then the evanescent field can generate repulsive dipole force. Based on this principle blue detuned doughnut beam is applied through the bottom of the prism to induce enough atomic reflectivity at the prism wall and cold Rb atoms are collected by the prism [5].

\section{Experimental}

In order to generate $TEM_{01}$ beam we use two-beam interference in a Michaelson type interferometer. Figure 1(a) shows schematic diagram of the experimental setup. A single mode Gaussian beam comes from Ti-saphire laser (maximum 1 W optical power). This beam is split into two equal intensity components by a non-polarizing beam splitter. These two beams are then passed through two different lenses $F_{1}$, $F_{2}$ in order to focus along its path. Besides, to recombine and make these beams interfere, two mirrors are placed in a Michelson interferometer configuration so the beams are reflected back to the beam splitter. One of the mirrors have PZT actuator mounted at its back so that mirror displacement control in nm level be achievable. Thus the beams again pass through the respective lenses and have further convergence. The focal lengths of lenses $F_{1}$ and $F_{2}$ are selected experimentally so that the beams are focussed close to the beam splitter after their second pass and their divergence beyond their respective focal points are appropriate to generate $TEM_{01}$ mode. In this case the effective focal lengths are \begin{math}f_{1}=22.2 \end{math} cm \begin{math}f_{2}=19.3 \end{math} cm. This results in minimum beam waists of 110$\mu$ and 96$\mu$ respectively at about 1.2 cm separation. Focal length of output collimator $F_{3}$ is 40 cm.

\begin{figure}[h]
\centering
\includegraphics[width=12cm]{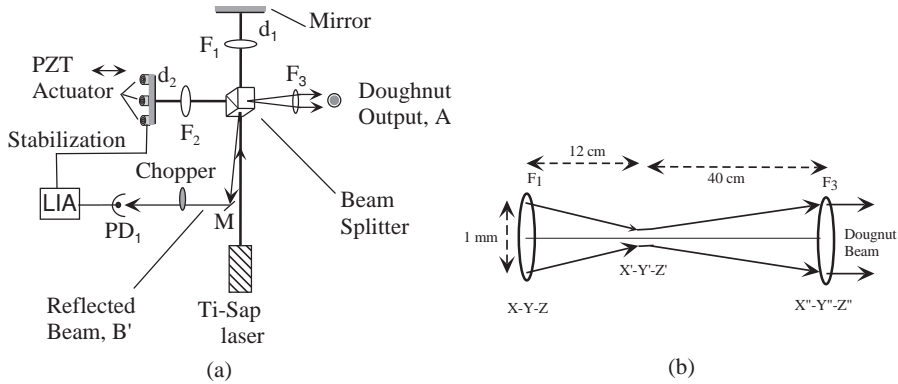}
\caption{Doughnut beam generation (a)Schematic diagram of optical layout for doughnut beam generation. LIA indicates lock in amplifier, M is a small miror. (b) Topological diagram of beam convergence and collimation.}
\label{fig:DoughnutBeamGenerationASchematicDiagramOfOpticalLayoutForDoughnutBeamGenerationLIAIndicatesLockInAmplifierMIsASmallMirorBTopologicalDiagramOfBeamConvergenceAndCollimation}
\end{figure}

\subsection{Analysis of beam mode formation}

Each of the beams in the interferometer can be analyzed by taking Fourier transform of the input beam. This is because Fourier transform can convert frequency domain spectrum into spatial distribution, which means a monochromatic collimated laser beam while passed through a converging lens results in a single point in focal plane where it forms the image. 

Although doughnut beam has been described by Laguerre polynomial, Gaussian function, angular momentum quantum number [13,14], however it has also been observed that focussed Gaussian beam acquires a axial phase [15] which is known as Gouy phase. With this we try to analyze the result with simplified theoretical model specific to our system and look into various characteristics experimentally.

Let $E_{a}(x,y,z)$ be initial electric field distribution (which is Gaussian in $X-Y$ plane) of the optical beam (propagating along Z-direction), figure 1(b). For analytical simplicity let's start with one of the two beams. While passing through converging lens let the electric field amplitude be $E(x',y',z')$ at focal plane $X'-Y'$  of lens $F_{1}$, With the help of Kirchhoff integral: 

\begin{eqnarray} E(x',y',z')=\int^{a}_{-a}\int^{a}_{-a}E_{a}(x,y,z)\frac{1}{r}\exp[jkr]\cos(\hat{k}, \hat{r})\rmd x\rmd y \label{kirch}
\end{eqnarray}


$\vec{r}$ position vector, $\vec{k}$  wave vector, the angle \begin{math}(\hat{k},\hat{r})\end{math}  being very small \begin{math} \cos(\hat{k},\hat{r})\approx 1\end{math} , $\lambda$  optical wavelength (780 nm), \begin{math} j=\sqrt{-1}\end{math}, initial beam waist as $a$ . In this situation \begin{math}\frac{1}{r}\end{math}  can be expressed as \begin{math}\frac{1}{(z'-z)} \end{math} , and

\begin{displaymath} \exp[jkr]\approx \exp\left[ jkz'+ \frac{(x'-x)^{2}}{2z'} + \frac{(y'-y)^{2}}{2z'} \right]\end{displaymath} 

\begin{displaymath} \fl \approx \exp \left[j\frac{\pi}{\lambda z'} \\ \left(x'^{2}+y'^{2}\right)\right]\exp \left[j \frac{2\pi}{\lambda} \\
\left( z'-\frac{2x'}{2z'}x-\frac{2y'}{2z'}y \right) \right] \\
\exp \left[j\frac{\pi}{\lambda z'}\left(x^{2}+y^{2}\right)\right]
\end{displaymath}

With Fraunhofer approximation, \begin{math}\exp \left[j \frac{\pi}{\lambda z'}\left(x^{2}+y^{2} \right)\right]\approx 1\end{math} , however we take this factor outside the integral rather than replacing by unity. Thus by parameterizing constant factor in the above integral as \begin{math}A_{1fi}\end{math} :

\begin{displaymath}
A_{1fi}=\frac{1}{j\lambda z_{i}}\exp[jkz_{i}] 
\exp\left[j\frac{\pi}{\lambda z_{i}}\left(x'^{2}+y'^{2}\right)\right]
\exp\left[j\frac{\pi}{\lambda z_{i}}\left(x^{2}+y^{2}\right) \right]
\end{displaymath}

with

\begin{displaymath}
A_{1f1}=\frac{1}{j\lambda f_{1}}\exp[jkf_{1}] 
\exp\left[j\frac{\pi}{\lambda f_{1}}\left(x'^{2}+y'^{2}\right)\right]
\exp\left[j\frac{\pi}{\lambda f_{1}}\left(x^{2}+y^{2}\right) \right]
\end{displaymath}

and

\begin{displaymath}
A_{1f2}=\frac{1}{j\lambda f_{2}}\exp[jkf_{2}] 
\exp\left[j\frac{\pi}{\lambda f_{2}}\left(x'^{2}+y'^{2}\right)\right]
\exp\left[j\frac{\pi}{\lambda f_{2}}\left(x^{2}+y^{2}\right) \right]
\end{displaymath}

If we look into the factor \begin{math}\exp\left[j\frac{\pi}{\lambda z_{i}}\left(x^{2}+y^{2}\right)\right]\end{math}, it is originated from initial beam state while \begin{math}\exp\left[j\frac{\pi}{\lambda z_{i}}\left(x'^{2}+y'^{2}\right)\right]\end{math}  is added due to the focal plane \begin{math}\left(X'-Y'\right)\end{math} , which if we compare to reference [15]  has some relation to Gouy phase. Now we get from equation ~\ref{kirch}, for \begin{math}i=1,2\end{math}

\begin{displaymath}
\fl E_{i}\left(x',y',z'\right)=A_{1f_{i}}\int^{a}_{-a}\int^{a}_{-a}E_{a} \\ \left(x,y,z\right)\exp\left[-2\pi j\left(\frac{x'}{\lambda f_{i}}x+\frac{y'}{\lambda f_{i}}y\right)\right]\rmd x\rmd y
\end{displaymath}

where \begin{math}f_{i}\end{math}  focal lengths of lenses. With  \begin{math}E_{a}\left(x,y,z\right)=A_{0}\exp\left[- \frac{x^{2}+y^{2}}{w_{0}^{2}}\right]\end{math} , where \begin{math}w_{0}\end{math}  is initial beam waist of input laser, \begin{math}A_{0}\end{math} its electric field amplitude \begin{math}E_{i}\left(x',y',z'\right)\end{math}  the integral can be written with the help of reference [16], as

\begin{eqnarray}
E_{i}\left(x',y',z'\right)=A_{1f_{i}} \frac{A_{0}}{2} \sqrt{\pi w_{0}^{2}} \frac{w_{0}^{2}}{2}
\exp \left[ - \left(\frac{\pi ^{2} w_{0}^{2}}{\lambda^{2} f_{i}^{2}}\right)  \left( x'^{2}+y'^{2} \right) \right]
\end{eqnarray}


This is Gaussian at \begin{math}\left(X'-Y'\right)_{i}\end{math}  planes along with a phase distribution function \begin{math}A_{1f_{i}}\end{math} . Its minimum beam waist as \begin{math}\frac{\lambda f_{i}}{\pi w_{0}}\equiv w_{0i}\end{math} . However beyond the focal plane the beam diverges due to its geometric structure, so the beam waist will linearly increase along its propagation. Let  \begin{math}\Psi_{i}\end{math} be half angle of divergence then each beam waist can be expressed as \begin{math}w_{i}=\left(w_{0i}+z'_{i}\tan \Psi_{i}\right)\end{math}. 

It has been observed that the two beams interfere after their respective focal planes and the topology of generated pattern does not change while the resulting beam diverges. Thus, while a third lens (\begin{math}F_{3}\end{math}) is used at some other distance apart in order to collimate the output beam with desired waist size it does not introduce any modification to beam pattern. It has also been observed that position of this third lens does not affect beam topology either. Let the lens \begin{math}F_{3}\end{math} be placed at \begin{math}X''-Y''\end{math} plane. The resulting beam diameter is 5 mm.

As the beam widths are finite and unequal so the interference pattern will depend not only on relative axial delay   between these two beams but also on its spatial optical field distribution. To simplify the analysis, let's introduce a phase part related to \begin{math}\left(\delta~ mod ~\lambda \right)\end{math} as \begin{math}\left(\sin\theta\right)\end{math}  factor to \begin{math}E_{2}\left(x',y',z'_{2}\right)\end{math} and \begin{math}\left(\cos\theta\right)\end{math} factor to \begin{math}E_{1}\left(x',y',z'\right)\end{math}, where \begin{math}\theta=\frac{2\pi}{\lambda}\left(\delta~ mod ~\lambda \right)\end{math} , then

\begin{displaymath}
\fl E\left(x'',y'',z''\right)=E_{1}\left(x',y',z'\right)\cos\theta +E_{2}\left(x',y',z'_{2}\right)\sin\theta
\end{displaymath}  

or

\begin{eqnarray}
\fl E\left(x'',y'',z''\right)=\sqrt{\pi}\frac{A_{0}w_{0}^{3}}{4} A_{1f_{1}}\exp \left[-\left(\frac{\left(x''^{2}+y''^{2}\right)}{\left(w_{01}+\left(z''_{2}+ \delta \right)\tan\Psi_{1}\right)^{2}}\right)\right]\cos\theta \nonumber\\ 
\lo+ \sqrt{\pi}\frac{A_{0}w_{0}^{3}}{4} A_{1f_{2}}\exp\left[-\left(\frac{\left(x''^{2}+y''^{2}\right)}{\left(w_{01}+z''_{2}\tan\Psi_{2}\right)^{2}}\right)\right]\sin\theta
\end{eqnarray}


This is the simplified generating function for different shape of the output beam for delay $\delta$.  This is called \begin{math}'\end{math}simplified$'$ because of a number of significant assumptions have been made so far. First and the prominent one is that the phase containing terms  \begin{math}A_{1f{i}}\end{math} are equal for i=1 and 2. Secondly, we implicitly assumed that each of the two interfering beams do not have any azimuthal phase or helicity to alter intensity distribution of output beams. This is true for Gaussian beam however we will see later that while the output beam interfere with its mirror image, it exhibits transverse phase distribution. The possible beam shapes that are obtainable experimentally is shown in figure 2,3. 


\begin{figure}[h]
\includegraphics[width=12cm]{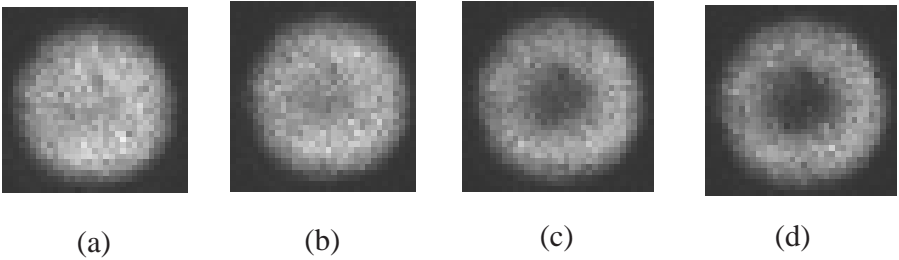}	
\caption{Various doughnut beam images at output A for various delay, (a) Taking this as reference, (b) 100 nm delay, (c) 250 nm delay, (d) 390 nm delay.}
\label{fig:VariousDoughnutBeamImagesAtOutputAForVariousDelayATakingThisAsReferenceB100NmDelayC250NmDelayD390NmDelay}
\end{figure}


\begin{figure}
\includegraphics[width=12cm]{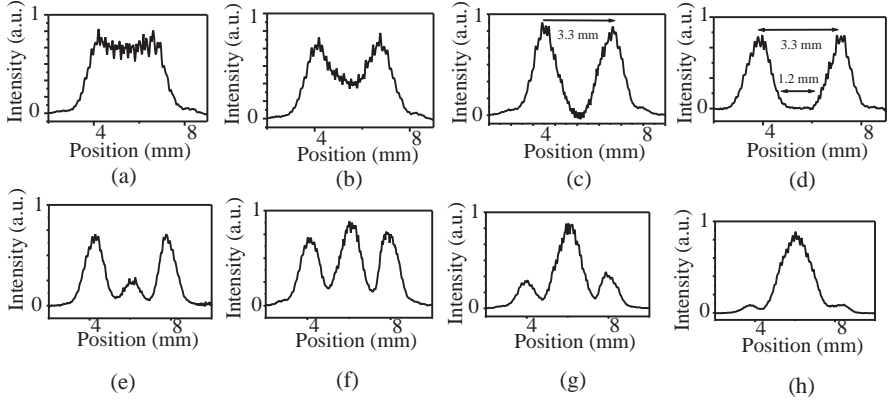}	
\caption{Beam profile for beams in A direction, (a)-(d) the delay correspond to that of figure 2(a)-(d), (e) 450 nm, (f) 550 nm, (g) 650 nm, (h) 750 nm delay.}
\label{fig:BeamProfileForBeamsInADirectionADTheDelayCorrespondToThatOfFigure2ADE450NmF550NmG650NmH750NmDelay}
\end{figure}

As seen in the figure 2\&3 not all the possible configurations correspond to doughnut shape except for figures within the range of figure 2. As we see from the above expression (3) that, at \begin{math}x''=0=y''\end{math} or, at beam axis, the \begin{math}E\left(x'',y'',z''\right)\end{math} becomes zero only when \begin{math}\theta \approx 3\pi{/}4\end{math} or \begin{math}-\pi{/}4\end{math} . However in this situation the optical intensity at the off-axial points does not vanish identically. In order to get a general expression covering all off-axial point we expand the exponential terms and rearrange to get 

\begin{eqnarray}
\fl E\left(x'',y'',z''\right)=A'_{1}\left[L_{01}\left[\frac{t_{1}t_{2}}{t_{1}-t_{2}}\right]\exp\left[-\left(t_{1}-t_{2}\right)\right]+t_{1}t_{2}\left(\frac{1+t_{1}-t_{2}}{t_{1}-t_{2}}\right)-1\right]
\end{eqnarray}


where 
\\
\begin{displaymath} A'_{1}=\left[{A_{1f1}A_{0}\sqrt{\pi}E_{10}w_{0}^{3}}\right] {/}4 
\approx \left[{A_{1f_{2}}A_{0}\sqrt{\pi}E_{10}w_{0}^{3}}\right]{/}4
\end{displaymath}

\begin{displaymath}t_{1}=\frac{r''^{2}}{\left(w_{01}+\left(z''_{2}+\delta\right)\tan\Psi_{1}\right)^{2}}\end{displaymath}

\begin{displaymath} t_{2}=\frac{r''^{2}}{\left(w_{02}+\left(z''_{2}+\delta\right)\tan\Psi_{2}\right)^{2}}
\end{displaymath}

where \begin{math}t_{1}= 2.25 \times 10^{5} r''^{2}\end{math} and \begin{math}t_{2}= 9.21\times 10^{4} r''^{2}\end{math} both in meter$^{2}$ where \begin{math}r''^{2}=x''^{2}+y''^{2}\end{math}. Thus \begin{math}\frac{t_{1}t_{2}}{\left(t_{1}-t_{2}\right)}<1\end{math} for \begin{math}r''<2.5\end{math}mm and \begin{math}\frac{t_{1}t_{2}}{t_{1}+t_{2}}<1\end{math} for \begin{math}r''<3.9\end{math} mm. We can verify that in equation (4) while \begin{math}x''=0=y''\end{math}, \begin{math}E\left(x'',y'',z''\right)=0\end{math} for all values of \begin{math}z''\end{math} and are vanishingly small at the beam edge. Furthermore, the term containing exponential factor increases faster with \begin{math}r''\end{math} than the other two terms, so it can further be approximated as

\begin{eqnarray}
E\left(x'',y'',z''\right)=A'_{1}\left[L_{01}\left[\frac{t_{1}t_{2}}{t_{1}-t_{2}}\right]\exp\left[-\left(t_{1}-t_{2}\right)\right]\right]
\end{eqnarray}


which is the expression for radial amplitude distribution for first order LG beam (\begin{math}LG_{01}\end{math} mode). Thus, with an odd multiple of \begin{math}\pi\end{math} or \begin{math}\left(2n+1\right)\pi\end{math} axial phase difference between the two beams we get \begin{math}LG_{01}\end{math} mode. Similarly, when the relative axial phase difference is \begin{math}2n\pi\end{math} maximum brightness at beam axis occurs with

\begin{eqnarray}
\fl	E_{b}\left(x'',y'',z''\right)=A'_{1}\left[\left[1+\frac{t_{1}t_{2}}{t_{1}+t_{2}}\right]\exp\left[-\left(t_{1}+t_{2}\right)\right]+L_{01}\left[\frac{t_{1}t_{2}}{t_{1}+t_{2}}\right]\right]
\end{eqnarray}


The figure 3(d) and 3(h) correspond to additional phase delay, introduced by the PZT actuator of figure 1(a).

\subsection{Topological Symmetry}

Thus we see from equation (3) that the achievable beam modes repeat after each \begin{math}\lambda\end{math} delay. The different beam profiles thus generated have topological symmetry, meaning the transformation from one beam structure to another is continuous, reversible and forms a closed set. Figure 3 shows the full range of available beam modes at several discrete optical delay in the range of \begin{math}\delta=780\end{math} nm, and for any delay \begin{math}\delta\end{math} more than integral multiple of \begin{math}\lambda\end{math}, the \begin{math}\left(\delta~ mod ~\lambda\right)\end{math} parameter becomes effective and the generated beam mode falls within that shown in figure 3. Thus we can categorize the set (figure 3) as a topological space with a pair $(X, T)$ consisting of a set $X$ of collection of possible beam modes and a family $T$ of subsets of $X$ so that:
(a) an empty set \begin{math}\Phi \in X\end{math} and \begin{math}X\in T\end{math} 
(b) $T$ is closed under arbitrary union
(c) $T$ is closed under finite intersection.
Although topologically it is not allowed to pierce or join two ends of a structure because of possible ambiguity in regeneration and reversibility, however creating a dark or bright center, the set of beam profiles that can be regenerated are unambiguous and without any arbitration. This argument is valid for zero intensity points. In this we can introduce a blank screen as empty set (\begin{math}\Phi\end{math}) which while added or subtracted from any member does not introduce any change, in practice this corresponds to generating delay among the two beams so that a particular beam configuration can be maintained, which in other words means the beam mode locking. Embedding this theory into an ideal experimental situation it corresponds to zero delay to a stabilized configuration. The condition (b) states $T$ is closed under arbitrary union, which implies that with any possible delay the achievable beam profile will always lie within that shown in the figure 3 or expressible by equation (3) as the effective delay \begin{math}\left(\delta~ mod ~\lambda\right)\end{math} always lies within $0$ to $780$ nm. Thus any subset of $X$ are open and the collection $T$ of open subsets can thus be called topology for $X$. 

With this categorical simplification we can further classify this experiment and its related results to look into its further characteristics. To proceed we broadly devide the set $X$ into two subsets, one with darker axial profile, figure 2 or figure 3(a)-3(d), and the other with brighter axial intensity distribution, figure 3(e)-3(h). Let these be subset $A$ and $B$. As we know that in the Michaelson interferometer one can get two beams, one goes away from the laser and other return towards the laser. We identify this second beam with a prime e.g. \begin{math}B'\end{math} for beams with brighter axial profile like that of figure 3(e) to figure 3(h) that returns towards the laser. We select $A$ and \begin{math}B'\end{math}, as these are simultaneously observable complementary beam structures the sum of which always gives the total beam intensity (assuming zero loss through the optics). Similarly $B$ and \begin{math}A'\end{math} can form another complementary set. Thus the delay that results in evolution of $A$ simultaneously generates \begin{math}B'\end{math}. To look into this situation analytically we take the following logical and analytical method.  

A homeomorphism between spaces $A$ and \begin{math}B'\end{math} is a continuous map \begin{math}f:A\rightarrow B'\end{math} which has an inverse continuous map \begin{math}g:B'\rightarrow A\end{math}, so that \begin{math}f(g(b'))=b'\end{math}  for all \begin{math}b'\in B'\end{math}, and \begin{math}g(f(a))=a\end{math} for all \begin{math}a\in A\end{math} . Thus the maps $f$ and $g$ introduces one-to-one and onto function (bijection) between the spaces $A$ and \begin{math}B'\end{math}. We can see from experiment that $A$ and \begin{math}B'\end{math} are simultaneously generated and forms two subsets of topological spaces that are distinctly measurable and identifiable in real space. In the following we will look into explicit expression of $f$ and $g$ in accordance to above formalism and investigate how this leads to interrelating the two spaces $A$ and \begin{math}B'\end{math} to stabilize $A$ using information from \begin{math}B'\end{math}.

We formulate the $f$ and $g$ in terms of delay and represent with \begin{math}\theta\end{math}, which is a function of \begin{math}\delta\end{math}. For $f$ and $g$ as continuous and inverse of each other it is sufficient to show that the square matrix representing $f$ and $g$ are inverse of each other. Let 
\begin{math} f=\left[\begin{array}{cc} \cos\theta & -\sin\theta \\ \sin\theta & \cos\theta 
\end{array}\right]
\end{math}

and by Gauss Jordan method 
\begin{displaymath}
g=
\left[\begin{array}{cc} \cos\theta & \sin\theta \\ -\sin\theta & \cos\theta 
\end{array}\right]
\end{displaymath}

thus \begin{math}g=f^{-1}\end{math}. Now rewriting expression (3) as

\begin{math}E\left(x'',y'',z''\right)=\left[p_{1}\cos\theta+q_{1}\sin\theta\right]
\end{math}

 where 
  
\begin{displaymath}
p_{1}=A'_{1}\exp\left[-\frac{\left(x''^{2}+y''^{2}\right)}{\left(w_{01}+\left(z''_{2}+\delta\right)\tan\Psi_{2}\right)^{2}}\right]
\end{displaymath}

and 

\begin{displaymath}
q_{1}=A'_{1}\exp\left[-\frac{\left(x''^{2}+y''^{2}\right)}{\left(w_{02}+\left(z''_{2}+\delta\right)\tan\Psi_{2}\right)^{2}}\right]
\end{displaymath}

represents the same physical situation in a convenient form. We can consider \begin{math}p_{1}\end{math} and \begin{math}q_{1}\end{math} as bases of the topology because any member of $T$ can be expressed as a linear combination of \begin{math}p_{1}\end{math} and \begin{math}q_{1}\end{math}. Taking \begin{math}E\left(x'',y'',z''\right)=a\end{math} and \begin{math}b'=E'\left(x'',y'',z''\right)\end{math} say, a matrix equation can be formed as follows

\begin{eqnarray}
\left[\begin{array}{c} a \\ b' 
\end{array}\right]=g\left[\begin{array}{c} p_{1} \\ q_{1} 
\end{array}\right]=\left[\begin{array}{cc} \cos\theta & \sin\theta \\ -\sin\theta & \cos\theta 
\end{array}\right]\left[\begin{array}{c} p_{1} \\ q_{1} 
\end{array}\right]
\end{eqnarray}


Or \begin{math}a=\left[p_{1}\cos\theta + q_{1}\sin\theta\right]\end{math} and \begin{math}b'=\left[p_{1}\cos\theta - q_{1}\sin\theta\right]\end{math}. In practice the negative sign in \begin{math}b'\end{math} arises as \begin{math}\theta\end{math} should change by $\pi$ for the beam \begin{math}B'\end{math}.
Thus the transformation relations corresponding to $f$ and $g$ are unitary, reversible and the elements that are generated in this way preserves total optical energy. Consequently, as is evident, for the delay (or \begin{math}\theta\end{math}) as independent variable of the transformation, the beam intensity corresponding to $a$ and \begin{math}b'\end{math}  becomes unequal, except for \begin{math}\theta\neq \left(4\pi+1\right)\pi{/}4\end{math}, which implies one of two beams acquires higher intensity in expense of the other. Table 1 shows the optical power of doughnut beam at various stages of beam mode. 

A point to be noted is that, although $a$ and \begin{math}b'\end{math} represents members of $A$ and \begin{math}B'\end{math} we can not use them in the above expressions to analyze their nature of continuity, that is test for monotonously increasing or decreasing nature will fail. This is because, as has been mentioned earlier, both $a$ and \begin{math}b'\end{math} are not uniform or their identity at different $\theta$ are not similar because of topological modifications. Thus, one can not get condition for maxima or minima by equating first derivative to zero.


\section{Results}
\subsection{Doughnut beam modes}

It has been observed that the output power of the doughnut beam varies with mode of the output beam. Table 1 shows the variation of beam mode and their respective optical power measured. Table 1 shows the optical beam intensity at various mode of the output beam [17, 18].

\begin{table}
\caption{\label{table}Optical power of doughnut beam $A$ at various beam modes}
\begin{tabular*}{\textwidth}{@{}l*{15}{@{\extracolsep{0pt plus
12pt}}l}}
\br
Laser power (mW)&Beam mode&Output ($A$) power&\begin{math}a : b'\end{math}\\
\mr

250&Figure 2(a),3(a)&145&2.3:1\\
   &Figure 2(c),3(c)&125&1.5:1\\
   &Figure 2(d),3(d)&105&1.0:1\\
\mr
610&Figure 2(a),3(a)&360&2.3:1\\
   &Figure 2(c),3(c)&300&1.4:1\\
   &Figure 2(d),3(d)&260&1.0:1\\
\mr
900&Figure 2(a),3(a)&620&2.2:1\\
   &Figure 2(c),3(c)&460&1.5:1\\
   &Figure 2(d),3(d)&380&1.0:1\\

\br
\end{tabular*}
\end{table}

Note that there is some loss of light at the optics, however the measured \begin{math}A:B'\end{math} beam power ratio remains constant.

The figure 1 shows that the optical delay is controlled by piezo-electric actuator attached to one of Michaelson-mirrors, it controls the beam mode. Besides, there may be electrical noise, external mechanical vibration as well as some other disturbance may create unsteady situation unless and otherwise special precaution is adopted. We have already observed that the set of beam modes form a topological set where $A$ and \begin{math}B'\end{math} are interrelated (7) and the null element of which while added or subtracted from the set $X$ does not have any effect. Now we shall use this idea to create a single element set or a stabilized doughnut beam output in which the null element will be functionally generated through locking circuit. This is useful in understanding locking the beam mode to a predetermined form. We know that $f$ and $g$ are inverse operations of each other. If there is some noise that arbitrarily generates $f$  a complementary process $g$ should be created to bring the system back to its initial state. We achieve this with the help of a photo-detector (PD$_{1}$) and lock-in amplifier combination and feed this signal back to the PZT actuator (figure 1(a)) so that the feed back signal can stabilize the system. 

This has been achieved quite simply because the \begin{math}B'\end{math} beam diameter is about 5 mm. With a 0.5 mm at the PD$_{1}$ window aperture of PD$_{1}$ while placed in its path, can discriminate a minute change in beam mode. Chopping the beam \begin{math}B'\end{math} and feeding the electrical output of the PD$_{1}$ to a lock in amplifier (LIA) the $g$ is generated. This is more evident from figure 2 that beam intensity distribution continuously varies with radial distance from beam axis. Thus while a change of beam profile with rise in $\theta$ can lead to increased PD signal, a corresponding fall in $\theta$ leads a change in PD signal on the reverse. 


\subsection{Focussed doughnut beam} 

One of several advantages that this beam have is its long propagation length; more than 3 meters observed in the laboratory. While doughnut beam is used to guide atom or as an instrument to manipulate atom in a localized space, the characteristics of such focussed beam will be useful. To observe doughnut beam mode under focus we used gold coated flat-top aperture fiber probe (figure 4(a)) to pick up optical signal from minimum beam waist.


\begin{figure}

\includegraphics[width=12cm]{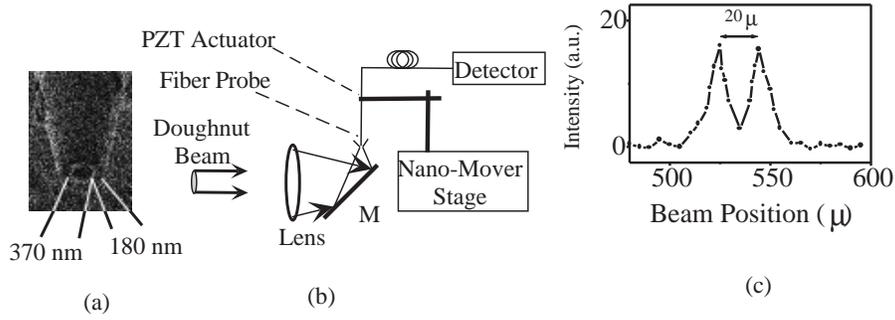}
\caption{Focussed doughnut beam (a)Flat-top aperture fiber probe as sampling window, (b) Measurement setup for minimum beam waist of free space focussed doughnut beam (\begin{math}f =8 \end{math}cm), (c) Doughnut beam profile measured at free space minimum beam waist by the probe.}
\label{fig:FocussedDoughnutBeamAFlatTopApertureFiberProbeAsSamplingWindowBMeasurementSetupForMinimumBeamWaistOfFreeSpaceFocussedDoughnutBeamF8CmCDoughnutBeamProfileMeasuredAtFreeSpaceMinimumBeamWaistByTheProbe}
\end{figure}

Figure 4(b) shows the schematic diagram of free space measurement setup. The fiber probe is mounted on a 3-axis nanomover stage (Melles Griot Inc.) in combination with a PZT actuator of 60 nm/Volt sensitivity. Figure 4(c) shows the measured free space beam profile. Accuracy of free space measurement has some limitation due to scattering at the fiber probe as well as non-zero transmission of the gold coating that covers the probe tip. Single tapered fiber probe has been fabricated by conventional chemical etching process and making a 180 nm gold coating over the tip. A ~370 nm aperture has been created by focussed ion beam (FIB) technique, figure 4(a). A details of such method can be found in [19].


\begin{figure}

\includegraphics[width=12cm]{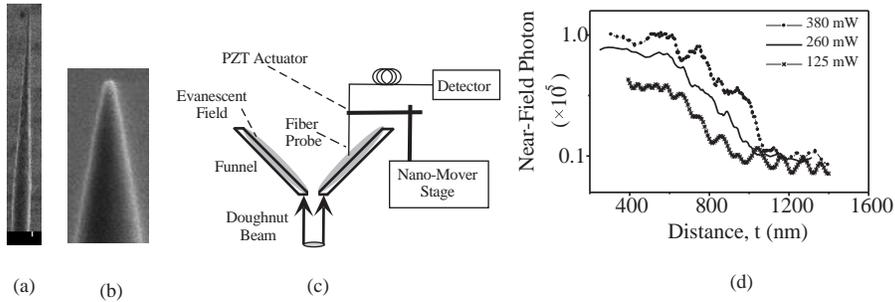}
\caption{Evanescent field, (a, b) protruded fiber probe used to detect optical near-field, (c)collimated doughnut beam in funnel-shaped evanescent setup, (d) evanescent field profile normal to the surface.}
\label{fig:EvanescentFieldAFocussedDoughnutBeamEvanescentSetupBProtrudedFiberProbeUsedToDetectOpticalNearFieldCEvanescentFieldProfile}
\end{figure}


After that we carried out the optical near-field measurement where the optical field is "blocked" beyond the probe tip and the scattering effect becomes minimum. Figure 5(c) shows the schematic diagram of the optical near-field measurement setup. Figure 5(a,b) shows the probe used in this measurement. Figure 5(d) shows the optical near-field distribution of the focussed doughnut beam measured in this way. About 20$\mu$ dark diameter has been recorded. It shows a dark hole probably exists along the minimum beam waist at the focal plane and the TEM\begin{math}_{01}\end{math} mode can have a continuous dark core even though the focal plane, when it is focussed. Such a characteristic can be useful while transporting cold atoms through the dark core and manipulating in a small dimension.

\subsection{Evanescent optical near-field}

The focussed doughnut beam profile as discussed above, has been the result of optical nearfield measurement. So at this stage we will discuss about optical near-field generated by doughnut beam in atom funnel system, measure the evanescent field and estimate most probable behaviour of Rb atoms at the funnel wall.

\subsubsection{Evanescent atom funnel}

In order to measure the optical near-field generated at the funnel surface of the atom funnel experiment, we use bare long fiber probe, figure 5(a), (b). A similar chemical etching technique, as described before, has been used with longer etching time to create long tip fiber probe for evanescent field measurement in funnel. Fiber probe-tip angle is about 23 degrees whereas for a short probe tip length of 6$\mu$ at fiber core of diameter ~40$\mu$. Thus in the evanescent field measurement over hollow prism it is inconvenient to use short fiber probe (for funnel surface, as the vertical wall angle is about 45\begin{math}^0\end{math}). As discussed earlier total internal reflection generates surface wave or evanescent field. The doughnut beam while is incident in such a way that a total internal reflection takes place, then a doughnut shaped evanescent field will be created at the prism surface. The measured optical near-field intensity distribution is as shown in figure 5(d).

The fiber probe is placed within the evanescent field range and it is kept in a predetermined position by applying a small vibration to the probe tip and collecting corresponding error signal. 

Application of doughnut beam in atom funnel has been developing [5]. In this system the necessary beam profile should have at least 250$\mu$ dark diameter along with 2.5 mm beam waist. Cold atoms are generated in magneto-optical trap and collected by the hollow prism with the help of the evanescent field created by doughnut beam. The prism wall thickness is about 3 mm and is composed  of three \begin{math}2.5\times 2.5\end{math} cm$^{2}$  glass plates glued together. The lower tip of the prism is polished to a flat triangular surface, at the center there is a 250$\mu$ hole. Doughnut beam is aligned at the bottom of the prism so that dark center lies at the prism hole and the beam gets total internal reflection at the inner surface. This illumination is synchronized with release of cold atoms from MOT so that the atoms bounce over the inner wall of the prism and collected through the bottom hole 

At three different  optical power evanescent field of various strength have been measured. However the measured range of the field primarily depend on the detector noise level and collection efficiency of the fiber probe. Thus with intense doughnut beam we can detect longer range of the field. 

The characteristic decay length at which the evanescent field intensity decays to \begin{math}\rme^{-1}\end{math} to that at the glass surface has been estimated from experimental result. The start of the exponential decay has been taken as the maximum value for the evanescent field. The minimum of the exponential decay corresponds to NEP of the detector system. Thus the range of the evanescent field is the range from the maximum evanescent field to the NEP level, which has also been estimated from experimental results. 

The probe tip cross section has about 42 nm diameter. Here we can make another practical approximation. While the doughnut beam is incident from the bottom of prism, the three planar walls of the prism surfaces will be uniformly covered with optical near field . This is because the wall thickness is about 3.0 mm whereas each wall area is \begin{math}2.5\times 2.5\end{math} cm$^{2}$ and the angle that the three edges make with vertical is 45 degrees. Besides, width of the bright ring is about 2.5 mm. Combining all these with repeated total internal reflections both at the inner and outer surface of the walls we can approximate to an almost uniform distribution of optical near-field. Thus with 360 mW beam the areal intensity distribution will be 213.3 W/m$^{2}$. With 5.54\begin{math}\times10^{-15}\end{math} m$^{2}$ fiber probe area optical field will be collected and measured. Assuming 70\% collection efficiency at the probe tip, because of refractrive index mismatch between probe and air, and also there be another 30\% optical power loss at fiber-photodetector (PD$_{2}$) coupling. To calibrate optical field distribution inside the funnel wall we take the noise equivalent power (NEP) of the system as the reference. Observed LIA noise level is 1.2$\mu$V. With a 10$^{9}$ gain of PD$_{2}$ gain circuit, it corresponds to 4.1\begin{math}\times\end{math}10$^{-2}$ pA current generated at PD$_{2}$. With a 50\% photo-voltaic conversion, this corresponds to 2.4\begin{math}\times\end{math}10$^{-2}$ pW optical power at the PD$_{2}$, (although the NEP level of the PD$_{2}$, at 200 Hz chopper frequency is ~1.4\begin{math}\times\end{math}10$^{-4}$ pW). This correspond to 4.1\begin{math}\times\end{math}10$^{-2}$ pW evanescent light at the fiber probe tip. At 780 nm wavelength the photon energy is 1.59 eV, and thus the NEP corresponds to 9.4\begin{math}\times\end{math}10$^{4}$ evanescent photons floating near the fiber probe tip. This is at a distance of 156 nm from the prism wall and at 360 mW doughnut beam power. 
Natural Rb atomic radius is 2.98$\r{A}$, so its cross sectional area is 2.8\begin{math}\times\end{math}10$^{-19}$ m$^2$. Therefore, while Rb atom approach the prism wall normally the number of photons it come across is 5.0\begin{math}\times\end{math}10$^{-5}$ times the number of photons faced by the detector probe. 


\subsubsection{Dipole force}

Grazing incidence reflection of atoms by evanescent field has been discussed since 1987 [20,21]. A blue detuned laser light while gets total internal reflection in glass gives a layer of repulsive evanescent field over the glass-air interface. At the blue detuning of laser line, because of faster change in laser field compared to that of atom it responds in a time frame lagging with respect to the external field. We can assume a frequency range (seen by the atom, or in atomic frame) in which this dipole force is appreciably high, lies close to resonance frequency of Rb atom. It is obvious that it can not happen at all higher frequencies or in any range of blue detuning. The principal difference between photon scattering (or photon reflection by atom) and this dipole force is that in the former case the incident photon frequency remains almost same to that of the scattered one, whereas in the latter its wave nature takes precedence. Thus without going into quantum mechanical analysis using dressed state formalism as well as drawing a sharp boundary between these two interaction mechanisms we can treat this atom kinetics semi-classically.

In free space if we assume a stationary photon distribution that corresponds to total doughnut beam power then the Rb atom at 10$\mu$K temperature will come across about 250 photons per second on its path, which in this case will be hypothetical limiting photon collision rate with cold and falling Rb atoms. To estimate the number of photons faced by Rb atoms at the prism wall, we take area under the "decay curve". However one may argue that when Rb atom comes close to the glass surface a localized field (dielectric constant) will change and a large number of photons will tunnel across the barrier to collide with the Rb atoms. We eliminate this situation for ground state Rb atoms having low multipole moment, falling with small velocity normal to the surface, thus repulsive dipole force will be stronger before reaching the photon tunneling range. Moreover, if Rb atoms do have sufficient velocity to reach to the photon tunneling range, the atoms will again be reflected due to velocity dissipation and photon scattering. Thus not only very slow atoms can reflect from the surface but also at higher atomic velocity the atom reflection can again be observed.


The optical dipole potential on atom can be expressed in terms of detuning $\Delta$ as, \begin{math}U_{0}=\hbar\Delta\sigma{/}2\end{math}, the factor \begin{math}\sigma\approx \left(\Gamma{/}2\Delta\right)^{2}TI{/}I_{0}\end{math}  incorporates laser intensity $I$, saturation intensity of Rb atom \begin{math}I_{0}=1.65 \end{math} mWcm$^{-2}$, its natural linewidth \begin{math}\Gamma=37.7\end{math} MHz, and a factor depending on polarization of the incident beam T\begin{math}\approx 5.7 \left( 2.57 \right) \end{math} for TM(TE) polarization [22] and \begin{math}U=U_{0}\exp\left[-2\kappa t\right] \end{math} where \begin{math}\kappa =\frac{2\pi}{\lambda}n^{2}\sqrt{\sin^{2}\alpha -1}\end{math}; $n$ is refractive index of glass (with respect to air, 1.45) and $\alpha$ angle of incidence \begin{math}{~}45^{0}\end{math}. However, instead of using theoretical value for $2\kappa$, we shall estimate it from experimental result. Thus the retarding dipole potential is distributed functionally similar to optical field intensity distribution. With 0.36 W doughnut beam power, the dipole potential distribution can be written as \begin{math}U_{dip}=5.75 \exp\left(-2\kappa t\right)\end{math} neV, for \begin{math}\Delta =1.5 \end{math}GHz, where $t$ is distance measured normal to the surface.


Apart from atom optical interaction described above there will be attractive van der Waals force that acts on Rb atoms due to proximity of the glass surface,\begin{math}U_{Vdw}(t)=-\frac{3\left(n^{2}-1\right)}{2.68\left(n^{2}+1\right)}\frac{\hbar\Gamma}{\left(k_{0}t\right)^{3}}\end{math}, [5] where \begin{math}k_{0}=\frac{2\pi}{\lambda} m^{-1}\end{math}. It has been known that atoms at a distance of more than \begin{math}0.2\lambda\end{math} the effective force is the repulsive dipole force and Van der Waals force have little effect on atom repulsion beyond such a distance [12]. Thus considering the reflected atoms we can neglect the Van der Waals force at a distance longer than 156 nm. So the repulsive optical force can be written as 

\begin{math}U_{opt}\approx 5.75\exp\left(-2\kappa t\right)\end{math} neV, for \begin{math}t>\end{math}156 nm. We know from evanescent field measurement that \begin{math}1{/}2\kappa =600\end{math} nm. By integrating this potential from infinity to 156 nm one can get work done on Rb atom while it moves toward the evanescent surface. Cold Rb atoms at 10$\mu$K temperature have most probable kinetic energy of 1.3 neV, which is smaller than the 5.75 neV. Thus cold Rb atoms will be reflected by the evanescent field created by the doughnut beam at the funnel surface.

\subsubsection{Dissipation}

In addition to dipole kinetics near the evanescent surface, atom may also have dissipation of its energy at the evanescent field interaction [23]. Now we will consider dissipative effect of evanescent field interaction in which atom looses part of its velocity near the surface. Rb atoms (say $^{85}$Rb) have two ground levels F=2,3 separated by 3.036 GHz frequency. It has been observed that while blue detuning increases from 1 GHz to 1.5 GHz the number of reflected atoms also increases [24]. With this trend we can assume that due to atomic velocity and the consequent Doppler shift a transition takes place from F=2 state by absorbing the photon (frequency $\nu_{1}$) and a subsequent emission at frequency $\nu_{2}$ to bring the atom back to F=2 state. Thus as the blue detuning increases so the transition probability rises. So from conservation of momentum, \begin{math}mu_{2}-mu_{1}=\left(\nu_{1}-\nu_{2}\right)h{/}c\end{math}, where $u_{1,2}$ are initial and final velocities respectively, $u_{1}$ and $\nu_{2}$ are in the same direction. And from conservation of energy, \begin{math}E_{1}+h\nu_{1}+0.5mu_{1}^{2}=E_{2}+hv_{2}+0.5mu_{2}^{2}\end{math}. Combining these two  \begin{math}u_{2}-u_{1}=\frac{h}{m\lambda_{trans}}\end{math}, where $m$ is mass of Rb atom, and \begin{math}\lambda_{trans}\end{math} corresponds to blue detuning (with respect to 5S F = 3 state), which in turn implies higher the blue detuning higher will be the atom reflection from the prism wall, as has been observed. Additionally we are considering atoms in motion so the atom will see the laser line with Doppler shift, which is \begin{math}\Delta\lambda_{Dop}=\pm(u_{1}\lambda{/}c)\end{math}, we shall take negative sign while atom is moving towards light. So \begin{math}u_{2}-u_{1}=\frac{h}{m\lambda_{trans}(1-u_{1}{/}c)}\end{math} , which is the magnitude of velocity rise opposite to the direction of $u_{1}$. Numerically we get \begin{math}u_{2}-u_{1} =23 \end{math}nm/sec. It has been observed that evanescent field may act dissipatively to incoming cold Rb atoms [23].


\subsubsection{Radiation force}

In addition to the dipole interaction, van der Waals force and velocity dissipation, there will be photon scattering from atoms as the atoms follow Maxwell Boltzmann velocity distribution. Let Rb atomic radius be \begin{math}r_{0} =0.298 \end{math}nm. Then most probable number of atomic collision with an evanescent photon will be \begin{math}n=\pi r_{0}^{2}l\int^{156}_{l}s_{0}\exp\left(-2\kappa t\right)\rmd t\end{math}, where $l$ is range of evanescent field, \begin{math}s_{0}\end{math}=4.7; photon calibration constant. This can be stated in terms of the figure 5(d) as \begin{math}n=\pi r_{0}^{2}ls_{0}\times\end{math}(area under the plot). Numerically \begin{math}n=1575\end{math} photons. 
The mechanism of scattering or photon reflection by atom can be described simply as follows. Conservation of momentum gives \begin{math}m\left(u_{2}-u_{1}\right)=2h{/}\lambda\end{math} or \begin{math}u_{2}-u_{1}=12 \end{math}mm/sec for each head-on colision with one photon. The estimated temperature of Rb atoms falling on the prism is 10$\mu$K [5]. Accordingly, the most probable velocity of Rb atoms will be 44 mm/sec. Thus the atoms need 4 'pure' scattering with photon if there be no induced dipole force and no velocity dissipation. As we can see from experiment the probable number of atom-photon collision is more than 6 (rather it is estimated to be around 1575), so atom reflection probability is high at 360 mW doughnut beam power.

\section{Azimuthal phase}

A characteristic property doughnut beam is its helicity or axial phase distribution [13]. Moving along circumference of such a beam cross section one faces varying optical phase covered by the path (or line integral of azimuthal phase), which can be expressed with a factor \begin{math}\exp\left(\rmi m\phi\right)\end{math} in its amplitude expression [7,13,14], where $m$ is azimuthal quantum number. Moreover, as we can see from Kirchhoff integral that the factor $A_{1fi}$ lies outside the integral and we keep it as constant factor for integration. However, considering \begin{math}\exp\left[\frac{j\pi}{\lambda f_{i}}\left(x^{2}+y^{2}\right)\right]\end{math} factor of $A_{1fi}$ we can say that with \begin{math}r'=\sqrt{x'^{2}+y'^{2}}\end{math} and \begin{math}r=\sqrt{x^{2}+y^{2}}\end{math}  of the beam its phase has additional parabolic distribution with its Gouy contribution [15]. Thus a radial phase  \begin{math}\phi_{r}=\frac{\pi}{\lambda f_{i}}\left(r'^{2}+r^{2}\right)\end{math} is not constant over minimum beam waist and its nature of variation remains unchanged while it diverges beyond the focal plane.

\begin{figure}
\includegraphics[width=12cm]{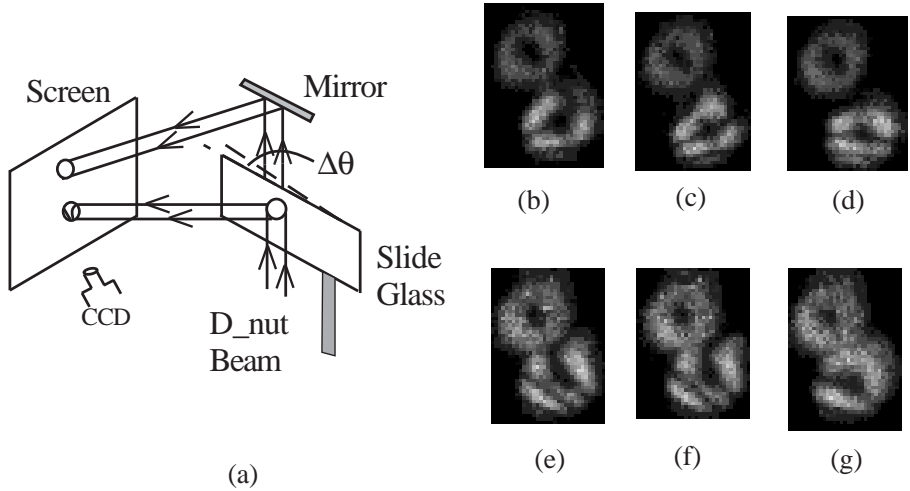}
\caption{Doughnut beam interference with its self image. (a) Optical setup, (b)-(f) various dark-bright intensity distribution observed by CCD camera. (g) This figure demonstrates presence of helical wave front while other phase contribution is minimum.}
\label{fig:DoughnutBeamInterferenceWithItsSelfImageAOpticalSetupBFVariousDarkBrightIntensityDistributionObservedByCCDCameraGThisFigureDemonstratesPresenceOfHelicalWaveFrontWhileOtherPhaseContributionIsMinimum}
\end{figure}

So the doughnut beam acquires both azimuthal [13] as well as radial phase distribution. To observe such a characteristic nature of the wavefront we performed interference of the doughnut beam with its self image. Figure 6(a) shows schematic diagram of such experimental setup. A 2.5 mm thick parallel face plane glass slide has been chosen for creating such a self image and interference by division of amplitude. As we can not modify separation of the faces of the glass slide so we adopt to change incidence angle (by several milli-radian) of the beam by rotating the slide in a vertical axis. At this stage we are not going to have quantitative analysis, so we will not discuss precise value of this angle and corresponding change in interference pattern. Thus a qualitative estimation will be sufficient for the discussion that follows.

It may be noted that the radial phase distribution that arises within Kirchhoff integral is inherent to each of the two beams. Thus it is independent of this doughnut beam generation and will remain present in all beam modes of figure 3. Now we try to establish the fact experimentally that such phase distribution indeed is present in the resulting beams described above. Both transmitted beam and reflected beams of figure 6(a) have been recorded simultaneously, as shown in figure 6(b) - 6(f). 

In figure 6(b)-6(f) we see that interference dark lines, arising due to parallel surface slide glass is linear in some part while non-linear at some other. Linear dark fringes arise when optical phase relationship between the two beams is linear, and similarly non-linear dark line correspond to non-linear phase relation. Due to helical wavefront and radial phase distribution, it is possible in the experiment of fig 6(a) that phase of a part of wavefront becomes linearly related while the other turns out to be non-linearly related. Thus, while incidence angle is changed (in milli-radian range) from 6(b) to 6(f), a transition from linear to non-linear dark fringe takes place. 

The transmitted part of the beam, going through the glass slide, does not show repetition of such pattern because of intensity mismatch of component beams. Furthermore, the azimuthal phase distribution is also demonstrated in figure 6(g), which shows only one 'spoke' [13] of the interference image that corresponds to helicity \begin{math}m=\frac{1}{2}\end{math}.  


\section{Discussion}

We see that the approximated analytical expression of the resultant beam of interference is followed by the experiments performed, however the effect of radial phase and azimuthal phase distribution that we did not take into account at doughnut beam formation stage, and it becomes evident through later part of experiment. In figure 2, 3 we see that the dark core of the doughnut beam does not arise only when condition \begin{math}\theta=\left(4n+1\right)\pi{/}4\end{math} is fulfilled, rather it tends to prevail through a range of delay. It may be due to radial as well as azimuthal and Gouy phase distribution of the two interfering beams dominate in such a zone which results in slow transformation of the dark axial region to bright zone.

In case of evanescent field and its role on atom reflection we looked into three possible mechanisms of atom photon interaction among these the dipole force and retardation effect of the optical near-field are polarization dependent while the scattering force does not depend on beam polarization. In addition dipole force is weaker than scattering force. Therefore, if we extend the analysis a bit further, it may be possible that the scattering force and the retardation effect have greater role to play in atom guidance of cold atoms through dark core laser beam.

Van der Waals force is the force in atomic range effective among atoms. When Rb atom comes closer to the glass wall its electric multipole moment and that of the glass surface will be responsible to generate such attractive force. The electric multipole moment will depend on atomic state, e.g. in the ground state the moment will be lower than that at excited state. However atom in the excited state while enter the evanescent region may subject to emit a photon in presence of dipole force field and get de-excited. As a result the atom may face a varying strength of Van der Walls force. However under the condition when Van der Waals force is weaker than dipole force and dipole force weaker than scattering force, the effect of such a localized and instantaneous effects may not have major effect in atom reflection in the atom funnel.

To transport cold atoms through its dark core it is necessary to have a suitable handle so that preferential force on such atom can be applied. A very weak but non-zero axial optical intensity profile can provide such a force through scattering. This in combination with radial optical potential field of the doughnut beam may transport and manipulate cold atoms more efficiently.

\section{Conclusions}

 Thus, we have generated tunable and stabilized doughnut beam through interference in a modified Michael interferometer. The optical power is sufficient to reflect cold Rb atoms, at around 10$\mu$K temperature. It has been observed that such a beam have radial as well as azimuthal phase distribution and helicity as \begin{math}\frac{1}{2}\end{math}. Out of various modes that available in this method, a doughnut mode have beam waist of ~2.5 mm and dark diameter more than 250$\mu$.

\end{document}